\documentclass[11pt,a4paper]{article}
\usepackage[utf8]{inputenc}
\usepackage[margin = 1.0cm]{geometry}
\usepackage{bm}
\usepackage{ marvosym }
\usepackage{ mathrsfs}
\usepackage{graphicx}
\usepackage{amsmath}
\usepackage{amssymb}
\usepackage{amsfonts}
\usepackage{float}
\usepackage[dvipsnames]{xcolor}
\usepackage{lineno}
\usepackage{adjustbox}
\usepackage{jcappub}
\usepackage{orcidlink}
\usepackage{placeins}

\newcommand{\apj}{ApJ}
\newcommand{\apjs}{ApJS}

\definecolor{srcolor}{rgb}{0.81, 0.09, 0.13}

\newcommand{\ukarcmin}[0]{$\mu$K-arcmin}
\DeclareMathOperator*{\argmax}{argmax}

\newcommand{\new}[1]{#1}

\title{Quantifying Bias due to non-Gaussian Foregrounds in an Optimal Reconstruction of CMB Lensing and Temperature Power Spectra}

\begin{document}

\author[1]{M.~Doohan\,\orcidlink{0000-0002-2195-0617}}

\author[2]{M.~Millea\,\orcidlink{0000-0001-7317-0551}}

\author[3]{S.~Raghunathan\,\orcidlink{0000-0003-1405-378X}}

\author[2]{F.~Ge\,\orcidlink{0000-0002-3833-8133}} 

\author[2]{L.~Knox}

\author[2]{K.~Prabhu}

\author[1]{C.~L.~Reichardt\,\orcidlink{0000-0003-2226-9169}}

\author[4,5]{W.~L.~K.~Wu\,\orcidlink{0000-0001-5411-6920}}

\affiliation[1]{School of Physics, University of Melbourne, Parkville, VIC 3010, Australia}
\affiliation[2]{Department of Physics \& Astronomy, University of California, One Shields Avenue, Davis, CA 95616, USA}
\affiliation[3]{Center for AstroPhysical Surveys, National Center for Supercomputing Applications, Urbana, IL, 61801, USA}
\affiliation[4]{Kavli Institute for Particle Astrophysics and Cosmology, Stanford University, 452 Lomita Mall, Stanford, CA, 94305, USA}
\affiliation[5]{SLAC National Accelerator Laboratory, 2575 Sand Hill Road, Menlo Park, CA, 94025, USA}

\date{\today} 

\abstract{ 
We estimate the magnitude of the bias due to non-Gaussian extragalactic foregrounds on the optimal reconstruction of the cosmic microwave background (CMB) lensing potential and temperature power spectra. The reconstruction is performed using a Bayesian inference method known as the marginal unbiased score expansion (MUSE). 
We apply  MUSE to a minimum variance combination of multifrequency maps drawn from the Agora publicly available  simulations of the lensed CMB and correlated extragalactic foreground emission.
\new{Taking noise levels appropriate to the SPT-3G D1 release, we find non-Gaussian foregrounds may bias the MUSE reconstruction of the lensing potential amplitude at the level of $(0.7\pm 0.3)\,\sigma$ when using modes up to $\ell_{max}=3500$.
We do not detect a statistically significant bias, finding a value of $(-0.4\pm 0.3)\,\sigma$, when restricted to lower angular multipoles, $\ell_{max}=3000$. }
This work is a first step toward understanding the impact of extragalactic foregrounds on optimal reconstructions of CMB temperature and lensing potential power spectra.

 }


\maketitle

\section{Introduction}

Measurements of the cosmic microwave background (CMB) have proven important to our understanding of cosmology. 
Motivated by this, major experimental programs seek to improve measurements of the CMB temperature and polarization anisotropies, such as the South Pole Telescope (SPT) \citep{bender_year_2018, prabhu_testing_2024}, Atacama Cosmology Telescope (ACT) \cite{mallaby-kay_atacama_2021}, Simons Observatory (SO) \citep{ade_simons_2019}, and the planned CMB-S4 experiment \citep{abazajian_cmb-s4_2016}. 
The ambitious science goals of these experiments motivates work to advance CMB data analysis techniques in order to fully exploit their science potential. 
In particular, achieving the stated science goals of CMB-S4 requires delensing the CMB power spectra to sharpen the acoustic peaks, in combination with precise measurements of the CMB lensing power spectra \cite{green_cmb_2017}.



\subsection{CMB Lensing}

As CMB photons travel to our telescopes from the last scattering surface at $z \sim $ 1100, they are deflected by distributions of matter along the line of sight, distorting our view of the primordial CMB. 
These distortions induce correlations between originally uncorrelated modes, and these correlations can be used to reconstruct the lensing deflection field \cite{lewis_weak_2006}. 
The net deflection is related to the gradient of the integrated potential along the line of sight, and thus the lensing deflection field yields information on the integrated gravitational potential from $z=0$ to the surface of last scattering. 
The maximum deflections of the CMB occur around $z\sim2$, with the CMB lensing kernel extending from $z\sim1$ to 4 \cite{lewis_weak_2006,abazajian_cmb-s4_2016}. 
CMB lensing therefore yields valuable information about the evolution of the universe and growth of structure at these redshifts, helping constrain the dark energy equation of state ($w$), the amplitude of mass fluctuations today ($\sigma_8$) and especially on the sum of the neutrino masses ($\Sigma m_{\nu}$) \cite{kaplinghat_determining_2003,lesgourgues_probing_2006,golshan_massive_2024}. 

While at very low noise levels the majority of lensing information will come from polarization  \cite{millea_optimal_2021}, CMB temperature anisotropies remain the most important information channel for reconstructing the lensing field in current wide surveys covering tens of percent of the sky \cite{sobrin_design_2022,ade_simons_2019}, given the higher noise levels on these wide field surveys in the pre-CMB-S4 epoch.
As these large sky areas are crucial to achieving the best cosmological constraints (except for searches for inflationary gravitational waves), temperature power spectra and the temperature map contribution to lensing reconstruction will remain extremely important for the next decade. 
%


\subsection{Foregrounds}
\label{sec:Foregrounds}
Complicating the study of CMB anisotropies, and their subsequent lensing by distributions of matter along the line of sight, is the emission of astrophysical foregrounds. 
Foregrounds can be split broadly into Galactic and extragalactic foregrounds. 
As this work focuses on improvements in determining lensing power by leveraging information at small scales in the lensed CMB, we consider only the extragalactic foregrounds: the Sunyaev-Zeldovich (SZ) effects, Cosmic Infrared Background (CIB) and radio galaxies. 

The SZ effects occur when CMB photons Compton scatter from free electrons within hot plasma \cite{sunyaev_microwave_1980}.
The effects are separated into the thermal SZ (tSZ), and kinematic SZ (kSZ) effects, where the former transfers energy to CMB photons due to the temperature of free electrons in the plasma, and the latter exchanges energy with the CMB photons due to the bulk motion of the electrons along the line of sight. 
The tSZ effect modifies the blackbody spectrum of the CMB, meaning knowledge of its spectral behaviour can be used in component separation techniques (see \S \ref{sec:ILC}).
The kSZ effect retains the blackbody spectrum of the CMB and so cannot be distinguished in this way, although it is sub-dominant compared to the tSZ. 

The CIB comes from dusty star-forming galaxies (DSFG), where UV and optical photons from the nascent stars warm the surrounding dust, producing thermal radiation \cite{lagache_sources_2005}. 
The brightest of these galaxies are usually masked down to some limiting flux, but there remains a background of unresolved CIB sources, whose spectral energy distribution (SED) can be modelled as a grey body.
The spatial distribution of the CIB is typically modelled as a Poisson distributed component and a clustered component \cite{casey_dusty_2014}.

Radio galaxies are powered by active galactic nuclei (AGN), which have varying spectral properties. Radio galaxies are separated into two categories: Steep spectrum radio sources where emission comes primarily from the dusty torus, occluding the black hole-accretion disk system, and flat spectrum sources, where emission comes primarily from the jets\cite{de_zotti_radio_2010}. As with the CIB, radio sources are masked down to a flux threshold, with the remaining unresolved sources typically modelled as a Poisson distributed component, with a power law frequency scaling (see e.g \cite{reichardt_improved_2021,balkenhol_measurement_2023}).

As the lensing of the CMB is detected through its coupling of modes in the Gaussian CMB, inducing non-Gaussian statistics, the presence of foregrounds is a confounding factor when inferring the power spectrum of the line of sight potential, due to the non-Gaussian spatial distribution of the foregrounds. Furthermore, as the extragalactic foregrounds come from areas of matter overdensity, there is a correlation between the line-of-sight potential ($\phi$), and the foregrounds.




\subsection{Lensing Estimation}
\label{sec:Lensing_Estimation}


Traditionally, estimates of the lensing potential have been formed by taking quadratic combinations of the lensed fields \cite{hu_mass_2002}. 
This quadratic estimate (QE) is based on the first order expansion of the lensed CMB fields in terms of gradients of the lensing potential, and so, neglects the information contained in higher order terms of the lensing effect on the CMB fields.
This minimum variance combination of the fields\footnote{As noted by Ref.\cite{maniyar_quadratic_2021}, the Hu \& Okamoto quadratic estimator is not actually the optimal minimum variance quadratic estimator. 
This does not affect the following arguments of sub-optimality with respect to Bayesian estimators.} becomes sub-optimal for experimental noise levels at or below  $\sim 5$\ukarcmin \cite{millea_optimal_2021}. 

An alternative to the QE is a Bayesian estimator, which makes use of all the available information in the fields \cite{lewis_weak_2006}. 
An important first step to this approach was achieved by  Ref.\cite{hirata_analyzing_2003} in computing the maximum a posteriori (MAP) estimate of the potential integrated along the line of sight, $\phi$, later improved upon by Ref.\cite{carron_maximum_2017} to be robust in the presence of masking and anisotropic noise.
However, these approaches produce estimates of the unlensed primary CMB fields that do not have an interpretation in the Bayesian framework.

The first algorithm to compute the \textit{joint} MAP of the CMB temperature field and $\phi$ was carried out by Ref.\cite{anderes_bayesian_2015}, later extended to include polarization and the tensor to scalar ratio, $r$, by Ref.\cite{millea_bayesian_2019}. 
Maximizing the joint-posterior with respect to the CMB fields, $\phi$, and cosmological parameters, ($\theta$ in the notation employed here), is an important step to forming a Bayesian estimate of $\theta$, however the joint MAP does not give useful constraints on $\theta$ (see e.g the toy example beginning around Eqn (48) of Ref.\cite{millea_bayesian_2019}).
In order to form a Bayesian estimate of $\theta$, and propagate the uncertainty on the inferred $\theta$, Monte Carlo sampling methods have been employed to explore the lensing joint posterior. Ref.\cite{millea_bayesian_2020} use the full lensing joint posterior to infer estimates on $r$ while marginalizing over the primordial CMB fields and $\phi$. This sampling method was applied to polarization data to form a joint inference of a scaling parameter on the lensing potential ($A^{\phi}$) and lensing-like peak smoothing effects on the CMB power spectrum ($A_L$) \cite{millea_optimal_2021}.

A roadblock to extending this tactic to large datasets is that the lensing problem is a hierarchical Bayesian problem, in the sense that the cosmological parameters, $\theta$, control the distribution of intermediate (or latent) variables that we do not directly observe.
Consider the target, marginalized posterior which we wish to sample, $\mathcal{P}(\theta |d)$, in terms of the joint-posterior $\mathcal{P}(T, \phi, \theta | d)$:

\begin{equation}\label{marginalPost_from_jointPost}
    \mathcal{P}(\theta |d) = \int  \mathcal{P}(T, \phi, \theta | d) dT d\phi, 
\end{equation}
where $\theta$ is a vector containing the lensing potential and CMB temperature band powers, and any other cosmological parameters of interest or calibration parameters that take experimental systematics into account.
These control the distribution of $T$ (the unlensed CMB temperature field) and $\phi$ (the integrated line-of-sight potential), which are latent variables in this problem as they are not directly observed. The dimensionality of $T$ and $\phi$ are equal to the number of pixels, $N_{pix}$,  on the observed sky map. 
For Ref.\cite{millea_optimal_2021}, which covered approximately 100 square degrees of sky, ($N_{pix} \simeq 10^5$), the time taken for the QE is $\sim$ 10 minutes, whereas the Bayesian MCMC sampling chain took $\sim $ 5 hours on 4 GPU's\footnote{This analysis used polarization data, but it still serves to highlight the computational challenge of using a bayesian estimator to probe $\phi$}. 
Sampling such a high dimensional posterior, where the most probable field and parameter values occupy a vanishingly small fraction of the bulk posterior, is a major hurdle for Bayesian inference. 

A promising approximation to the integral in Eqn (\ref{marginalPost_from_jointPost}) which uses the joint-MAP, the Marginal Unbiased Score Expansion (MUSE), was proposed in Ref.\cite{millea_muse_2022}, and is discussed in \S\ref{sec:muse}.
Ref.\cite{qu_impact_2024} uses MUSE for polarization only simulations assuming CMB-S4 experimental conditions to infer the lensing potential power spectrum ($\theta=C_L^{\phi\phi}$), and includes the polarized foregrounds as a Gaussian contribution to the noise
Ref.\cite{ge_cosmology_2024} apply MUSE to 2 years of polarized data from SPT-3G, but ignore the foreground contribution as they estimate the maximum contamination to be two orders of magnitude lower than their noise levels.

While foreground bias, and mitigation techniques, have been explored for the QE using temperature data \cite{engelen_cmb_2014,osborne_extragalactic_2014,madhavacheril_mitigating_2018,schaan_foreground-immune_2019,baxter_dark_2019,sailer_lower_2020,raghunathan_cross-internal_2023,shen_auto_2024}, to our knowledge this has not been done with temperature for a Bayesian estimator like MUSE. 
To assess the level of bias with an optimal lensing estimator, we use realistic simulations of temperature data, with non-Gaussian spatial distribution of the foregrounds that are correlated with the line-of-sight potential, as input data.
For our analysis model, we assume a Gaussian distributed foreground model, and include it in the Bayesian inference of the power spectra of temperature, $\phi$, and the foregrounds (Sec \ref{sec:methodology_and_data}). 

\section{Methodology and Data}
\label{sec:methodology_and_data}

\subsection{Agora Simulations}  
\label{sec:Agora}

To test for potential biases to lensing estimators related to the non-Gaussianity of extragalactic foregrounds, we need simulated sky realizations including the lensed CMB and foregrounds. 
For this, we turn to the publicly available Agora simulation suite \citep{omori_agora_2024}. 
The Agora simulations include realizations of the unlensed CMB anisotropy consistent with a Planck 2013 model \citep{ade_planck_2014}. 
The CMB anisotropy is gravitationally lensed by a lensing convergence field drawn from the dark-matter only MultiDark-Planck2 (MDPL2) N-body simulation \cite{klypin_multidark_2016}. 
The lensing convergence field in Agora is calculated by ray-tracing the MDPL2 simulation up to $z=8.6$, and then adding a small additional Gaussian realization to match the expected contribution from higher redshifts. 
Thus we have three maps per sky realization: the unlensed primary CMB anisotropy, the lensing convergence field, and the lensed CMB anisotropy. 

The Agora simulations also provide foreground maps of radio galaxies, CIB, and tSZ and kSZ effects. 
All foreground components are drawn from the same dark matter halo catalogs in the MDPL2 simulations used to lens the CMB, and are lensed using the integrated lensing field up to the location of the source. 
The foreground simulations thus reflect the non-Gaussian distribution of foregrounds in the real world, as well as plausible correlations between the different signals.
Therefore we expect the Agora simulation maps to provide a reasonable estimate of the extent to which lensing estimators will be biased by the tSZ, kSZ, CIB and radio sky.
We refer the reader to \citet{omori_agora_2024} for a complete description of how the foreground signals are generated. 

Using the Agora simulations, we create data maps representative of those made with data from the SPT-3G instrument on the South Pole Telescope \cite{benson_spt-3g_2014,bender_year_2018,sobrin_design_2018}.
CMB analyses mask bright point sources and clusters in the maps. 
Reflecting this point, we mask regions of the Agora maps where there are radio and dusty galaxies with fluxes $\ge 4$\,mJy at 150GHz. 
Clusters in the maps are masked following Ref.\cite{raghunathan_assessing_2022}, with a mass threshold\footnote{Where $M_{500c}$ is the mass within a radial distance of a cluster that has a density 500 times that of the critical density of the universe.} of $M_{500c} = 3 \times 10^{14} M_{\odot}$.
This radio/CIB and cluster mask covers approximately 4\% of a map.
We add to the Agora maps Gaussian realizations of instrumental and atmospheric noise, deconvolved by an SPT-3G beam appropriate to each frequency band, where the white noise levels are representative of 2 years of SPT-3G data. The beam shapes are taken from Ref.\cite{spt-3g_collaboration_measurements_2021}, and are approximately Gaussian. The white noise levels, atmospheric noise parameters and beam FWHM are given in Table.~\ref{Table:map_dimensions_and_noise_levels}. 
These simulation maps with bright sources masked, and noise added are the input maps to \S\ref{sec:ILC}.

\subsection{Minimum Variance Maps}
\label{sec:ILC}

We combine the three individual frequency maps from Agora via a minimum variance internal linear combination (MV-ILC) \cite{tegmark_high_2003, eriksen_foreground_2004, cardoso_component_2008} into a single map to minimize foreground variance in the output map while preserving signal-to-noise, and reducing the computational complexity of modeling individual foreground fields. 
One might use a constrained ILC (cILC) map as in Ref.\cite{remazeilles_cmb_2011} to zero out a problematic foreground signal, such as the tSZ or CIB, at the cost of more noise variance in the final map. However, this can amplify a foreground signal which is not projected out,
whereas projecting out tSZ and CIB simultaneously increases the noise by factors of 2 to 3 in comparison to the MV-ILC because there are fewer degrees of freedom to use to minimize the variance \cite{raghunathan_cross-internal_2023}.
Furthermore, the CIB has the added complication of there being multiple populations of DSFG's, so any CIB signals that do not follow the chosen SED may in fact have their contribution boosted in the final map\cite{mccarthy_component-separated_2024}. Mitigating this by projecting out further moments of the CIB SED as in Ref.\cite{mccarthy_component-separated_2024} exacerbates the 2$\times$-3$\times$ noise cost of nulling tSZ and one CIB SED.

The cross-ILC proposed by Ref.\cite{raghunathan_cross-internal_2023} uses two cILC maps, a tSZ-free and CIB-free map, as inputs to the cleaned gradients method of Ref.\cite{madhavacheril_mitigating_2018}. An analogous approach may be attempted for a Bayesian lensing estimator, where the tSZ-free and CIB-free maps enter the data vector as two separate channels. This does increase the complexity as two foreground fields, with amplified tSZ(CIB) contributions in the CIB-free(tSZ-free) map, must be modeled, as well as doubling the computational load in solving Eqn (\ref{eqn:MAP_condition}). Given the increased noise in these methods with respect to the MV-ILC, and further complexity in modeling the residual foregrounds, we leave explorations of these alternate ILC techniques to future work.

\begin{figure}
    \centering
    \includegraphics[width=1\linewidth]{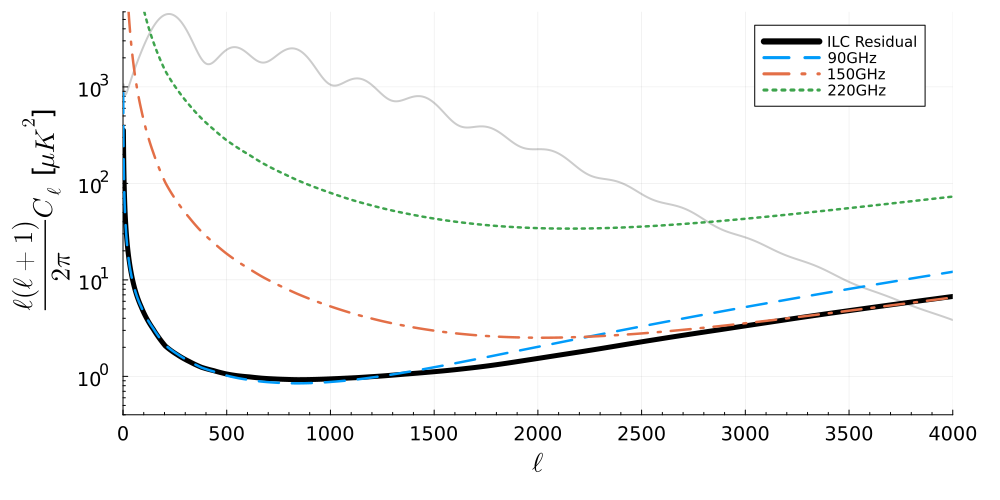}
    \caption{The total foreground and beam-deconvolved noise power at 90GHz (blue dashed), 150GHz (orange dot-dash), 220GHz (green dotted), and of the final ILC map (black solid). The CMB power spectrum of lensed temperature is shown in grey}
    \label{fig:ilc_sig_noise_power}
\end{figure}

Following Ref.\cite{delabrouille_full_2009}, we estimate the ILC weights from the power spectra of the full-sky Agora simulation maps. 
The weights ($\mathbf{w}_{b}$) used are:
\begin{equation}\label{eqn:mv-ilc_weights}
\mathbf{w}_{b} = \frac{\mathbf{a}^{\dagger}\mathbf{C}_{b}^{-1}}{\mathbf{a}^{\dagger}\mathbf{C}_{b}^{-1}\mathbf{a}}
\end{equation}
where $\mathbf{a}$ is a vector of ones in CMB thermodynamic units and $\mathbf{C}_{b}$ is the binned map to map covariance as a function of angular scale. This covariance can be expressed as:


\begin{equation}\label{ilc_covariance}
        \mathbf{C}_{b} = \begin{bmatrix}
        C_{b}^{90 \times 90} & C_{b}^{90 \times 150} & C_{b}^{90 \times 220}\\
        C_{b}^{150 \times 90} & C_{b}^{150 \times 150} & C_{b}^{150 \times 220}\\
        C_{b}^{220 \times 90} & C_{b}^{220 \times 150} & C_{b}^{220 \times 220}\\
        \end{bmatrix},
        C_b^{X \times Y} = \sum_{\ell \in b} \frac{2 \ell + 1}{4 \pi} C_{\ell}^{X \times Y}
\end{equation}
where the bins $b$ are chosen to be uniformly spaced at a width $\Delta \ell = 50$.
These weights are applied in spherical harmonic space to the noisy, bright-source-masked Agora simulated maps to produce a single full-sky, MV map. 
\new{Thirteen patches of area $\sim$ 1500 square degrees are cut from the full-sky ILC map (matching the SPT-3G Main field \cite{bender_year_2018,sobrin_design_2018,prabhu_testing_2024} in area), and projected to an oblique Lambert azimuthal equal-area projection (ProjZEA) with 2.25$^\prime$ square pixels.
These ProjZEA, minimum-variance ILC maps are referred to as the ILC maps hereafter.}
Figure \ref{fig:ilc_sig_noise_power} shows the contribution in power by foregrounds and noise at each frequency band, as well as the total ILC residual power, which includes the residual noise and non-CMB signal powers.

\begin{table}[h]
    \centering
    \caption{Dimensions of the ILC maps followed by white noise levels ($\Delta_T$), atmospheric noise level parameters, ($\ell_{\rm{knee}}, \alpha_{\rm{knee}}$) and beam full width at half maximum (FWHM) for the three input maps to the ILC. The area shown is the area after masking point sources and clusters. The effective white noise level and beam FWHM are given for the ILC map.}
    \begin{tabular}{ c|c c c c }
        \hline
        \hline
        Area [deg$^2$] & \multicolumn{4}{c}{1415}\\
        pixel width  & \multicolumn{4}{c}{$2.25^{\prime}$}\\
        \hline
        \hline
           & 95GHz & 150GHz & 220GHz & ILC \\ 
        $\Delta_T$ [\ukarcmin] & 4.74 & 3.95 & 14 & 4\\ 
        $\ell_{\rm{knee}}$ & 1200 & 2200 & 2300 & -\\
        $\alpha_{\rm{knee}}$ & -3 & -4 & -4 & -\\
        Beam FWHM & $1.7^{\prime}$ & $1.2^{\prime}$ & $1^{\prime}$ & $1.3^{\prime}$\\
        \hline
        \hline
    \end{tabular}
    \label{Table:map_dimensions_and_noise_levels}
\end{table}

\subsection{Lensing Model}
\label{sec:lensing_model}


\new{The three single-frequency, full-sky maps, from which the full-sky MV map is made, are source masked before taking the spherical harmonic transform to minimize ringing around the sources. 
In harmonic space, we apply a bandpass filter of the form $e^{-(50/\ell)^6} e^{-(\ell/4200)^{20}}$, and combine the maps using the weights from equation (\ref{eqn:mv-ilc_weights}) to form the MV map. 
After transforming back to map space, thirteen 1500\,deg$^2$ ILC maps (see \S\ref{sec:ILC}) are extracted from the full-sky MV map using a ProjZEA reprojection.}
This reprojection picks up two factors of the healpix map pixel window function (PWF), as well as the PWF of the 2.25$^{\prime}$ ProjZEA map. So the \new{effective} PWF is  $\mathbb{PWF}=\mathbb{PWF}_{\rm{ProjZEA}}\mathbb{PWF}_{\rm{Healpix}}^2$.
\new{Finally the ILC maps are apodised and a bandpass filter is applied in Fourier space.} The low-pass end of the Fourier filtering is allowed to vary to explore the effect of including smaller scales on the lensing estimate.

We model an ILC map as a combination of three fields: (i) $T$, the unlensed CMB temperature field, (ii) $\phi$, the integrated line of sight potential that lenses $T$, and (iii) $g$, the foreground contamination remaining after the ILC. We assume all three fields are isotropic and Gaussian distributed, and the following data model\footnote{As written in Eqn (\ref{eqn:data_model}), none of the masking/filtering is applied to the noise in this model. This is to avoid having to invert (implicitly) a masked noise covariance that is neither diagonal in pixel space nor harmonic space, which slows down the computation of the MAP of the fields in Eqn (\ref{eqn:MAP_condition}). Although trivially done for simulated data, a prescription for doing this with real data can be found in section 3.8 of \cite{millea_optimal_2021}. Furthermore, Ge et al \cite{ge_cosmology_2024} find little impact on their final results when using this approximation in their posterior model.}:

\begin{equation}\label{eqn:lens_model_field_distribution}
 T \sim \mathcal{N}(0, \mathbb{C}_{T}) \quad
 \phi \sim \mathcal{N}(0, \mathbb{C}_{\phi}) \quad
 g \sim \mathcal{N}(0, \mathbb{C}_{g})
\end{equation}

\begin{equation}\label{eqn:data_model}
    d' = \mathbb{M}_{\rm{Fourier}}\cdot\mathbb{M}_{\rm{apod}}\cdot\mathbb{PWF} \cdot\mathbb{M}_{\rm{Healpix}}\cdot\mathbb{M}_{\rm{source}}\cdot(\mathcal{L}(\phi)\cdot T + g) + n
\end{equation}
where $d'$ is the simulated data, $\mathbb{M}_{\rm{Fourier}}$ is a bandpass filter, $\mathbb{M}_{\rm{apod}}$ is an apodisation mask, $\mathbb{PWF}$ is the pixel window function described above,  $\mathbb{M}_{\rm{Healpix}}$ represents the anti-alias and DC filtering that was applied to the full-sky map for the ILC before projecting to ProjZEA, $\mathbb{M}_{\rm{source}}$ is the mask for sources and clusters, and $\mathcal{L}(\phi)$ is the lensing operation which we perform with \textsc{LenseFlow}\cite{millea_bayesian_2019}.
Note that in simulating data with Eqn(\ref{eqn:data_model}), we \new{ simulate the fields and apply the $\mathbb{M}_{\rm{Healpix}}$ and $\mathbb{PWF}$ operations in Fourier space, thus assuming the flat sky approximation and neglecting projection effects}. 

The covariances of the fields are diagonal operators in Fourier space,  controlled by bandpower amplitudes ($A_b^{XX}$), which scale the power spectra of the fields with respect to some fiducial model for all $\ell$ within bin $b$ as shown below. The joint-likelihood for our model largely follows \cite{millea_muse_2022} and \cite{millea_bayesian_2019}\footnote{We use the shorthand $\frac{x^2}{\mathbb{C}_x} \equiv x^{\dagger{}}\mathbb{C}_x^{-1}x$}.

\begin{equation}\label{eqn:scaled_fid_pwr_spec}
 \mathbb{C}_{X} = \rm{Diagonal}(A_b^{XX} C_{\ell, \rm{fid}}^{XX})
\end{equation}


\begin{equation}
 -2\rm{log}\mathcal{P}(d,T,\phi,g | A_b^{TT}, A_b^{\phi \phi }, A_b^{gg}) = \frac{(d - \mathbb{A}\cdot(\mathcal{L}(\phi)\cdot T + g))^2}{\mathbb{C}_n} + \frac{T^2}{\mathbb{C}_T(A_b^{TT})} + \frac{\phi^2}{\mathbb{C}_{\phi}(A_b^{\phi\phi})} + \frac{g^2}{\mathbb{C}_g(A_b^{gg})} \nonumber
 \end{equation}
 \begin{equation}
 + \rm{log}( | \mathbb{C}_T(A_b^{TT}) |) + \rm{log}( | \mathbb{C}_{\phi}(A_b^{\phi\phi}) |) + \rm{log}( | \mathbb{C}_{g}(A_b^{gg}) |) + \rm{log}( | \mathbb{C}_n |) - 2\rm{log}(\mathcal{P}(\phi)) \label{eqn:joint_likelihood}
\end{equation}
Where the operator $\mathbb{A}$ describes any masking/filtering applied to the map as in Eqn (\ref{eqn:data_model}), $d$ is the real or simulated data, $\mathbb{C}_n$ is the covariance of the noise in the map, and $\mathcal{P}(\phi)$ is the super sample prior. This last term was introduced by Ref.\cite{millea_muse_2022} when computing the MAP of a similar joint-likelihood with respect to the CMB and $\phi$ fields (see \S\ref{sec:muse}) to minimize a mean-field contribution to $\phi$.


\subsection{MUSE}
\label{sec:muse}

To perform the Bayesian inference necessary to optimally extract lensing information from our simulations, we use an approximation to performing the integral introduced in equation (\ref{marginalPost_from_jointPost}), known as the Marginal Unbiased Score Expansion, or MUSE \cite{millea_muse_2022}\footnote{\href{https://github.com/marius311/MuseInference.jl}{MuseInference.jl}}. 
As mentioned in \S\ref{sec:Lensing_Estimation}, the integral in equation (\ref{marginalPost_from_jointPost}) is intractable, given the dimensionality of the fields ($T$, $\phi$ and $g$) to be integrated over. 
For the experimental specifications adopted in this work, (see Tab.\ref{Table:map_dimensions_and_noise_levels}) the number of pixels on the $\sim$ 1500 square degrees of sky observed is $N_{\mathrm{pix}} \sim 10^6$ per field. 
The utility of MUSE is in reframing this high-dimensional integral as an optimization problem, with a dimensionality equal to the number of parameters being solved for ($\theta_i$). The chosen summary statistic that is optimized in MUSE
is a data and simulation combination of the score, the gradient with respect to parameters of the joint log-likelihood.
The first step to this approach is maximizing the joint likelihood with respect to the latent variables,  ( [$T$, $\phi$, $g$] in this work), known as the Maximum a Posteriori (MAP), i.e find $T_J$, $\phi_J$ and $g_J$ that satisfy:

\begin{equation}\label{eqn:MAP_condition}
T_J, \phi_J, g_J = \argmax_{T, \phi, g} [\mathrm{log}\mathcal{P}(d, T,  \phi, g|  \theta )]
\end{equation}
where the maximization is carried out in a manner similar to \cite{millea_bayesian_2019}, using a modified version of the code \href{https://github.com/marius311/CMBLensing.jl}{CMBLensing.jl}. 
With $T_J$, $\phi_J$, and $g_J$ in hand, the vector of parameter values, $\hat{\theta}$, that solve the score equation below are the output of the MUSE estimate.

$$
s^{\rm{MUSE}}_i(\theta,d) =
$$
\begin{equation}\label{eqn:muse_score_equation}
     \frac{d}{d \theta_i} \mathrm{log}\mathcal{P}(d, T_J, \phi_J, g_J | \theta ) - \Big{\langle} \frac{d}{d \theta_i} \mathrm{log}\mathcal{P}(d', T_J, \phi_J, g_J| \theta) \Big{\rangle}_{d' \sim \mathcal{P}(d'|\theta)} = 0
\end{equation}
where $d'$ are simulated data, generated at the parameter values $\theta$, and the derivatives are carried out with Automatic Differentiation (AD). 
The expectation value is calculated by averaging over N sets of simulated data. The vector of parameters that balances equation (\ref{eqn:muse_score_equation}) ($\hat{\theta}$) is obtained as follows...
\begin{enumerate}
  \item generating $N$ sets of data, $d'$ at a starting guess for the parameter values, $\theta_0$
  \item finding $T_J$, $\phi_J$, and $g_J$ that satisfy Eqn (\ref{eqn:MAP_condition}) for the real and simulated data
  \item Iterating over steps 1 and 2 to find the $\hat{\theta}$ that satisfies Eqn (\ref{eqn:muse_score_equation})
\end{enumerate}
Step 3 is achieved by using a pseudo Newton-Raphson method to approach $\hat{\theta}$ by updating it as 
\begin{equation}\label{eqn:theta_update}
\hat{\theta}_{i+1} = \hat{\theta}_{i} - K^{-1} s^{\rm{MUSE}}
\end{equation}
The $K$ matrix for the Newton-Raphson step would ideally be the Jacobian of $s^{\rm{MUSE}}$, but in practice we take the covariance of the summary statistic, (covariance of the term in the expectation value brackets in Eqn (\ref{eqn:muse_score_equation})), $J$ and set $K=-\rm{diag}(J)$.


Although one could calculate the covariance ($\Sigma$) of the estimate $\hat{\theta}$ via Monte-Carlo, MUSE allows for a less computationally intensive route to $\Sigma$. 
This requires $J$, and the average of the response of the summary statistic to perturbed values of the parameters, $H$.
The covariance of $\hat{\theta}$ is then given by\footnote{We refer the reader to Ref.\cite{millea_muse_2022} for details.} 

\begin{equation}\label{eqn:muse_covariance}
    \Sigma = H^{-1} J H^{-\dagger}
\end{equation}
Note that the final iteration to compute $\hat{\theta}$ means that a set of simulated scores is automatically available to obtain their sample covariance, which will be a good approximation to $J$ provided enough simulations are used. 
The only extra step to compute $\Sigma$ is computing $H$, either with Finite Difference methods (FD) or AD.

Finally we note that if a MUSE inference is carried out to find a set of parameters $\hat{\theta}$ = ($\theta^{\prime}, \theta_{\rm{nuisance}}$), then one can consider excising $\theta_{\rm{nuisance}}$ from $\hat{\theta}$ and $\Sigma$ as effectively marginalizing over $\theta_{\rm{nuisance}}$. This can be understood by assuming that the output $\hat{\theta}$ is the parameter vector that maximizes the \textit{marginalized} likelihood $\mathcal{P}_m(d|\theta)$, which is asymptotically true for MUSE in the limit of many simulations. If we treat $\theta_{\rm{nuisance}}$ as the latent parameters in an equation analogous to Eqn (\ref{eqn:muse_score_equation}), but in terms of $\mathcal{P}_m$, that score equation is automatically satisfied as $\theta_{\rm{nuisance}}$ already maximizes $\mathcal{P}_m(d,\theta_{\rm{nuisance}}|\theta^{\prime})$.


\subsection{Modeling the Residual power} 
\label{sec:modeling_the_residual_power}

As noted in \S\ref{sec:Foregrounds}, the key astrophysical emission signals are expected to be tSZ, CIB and radio point sources, all of which are included in the Agora simulations. 
The foreground signals are largest (relative to the CMB) in temperature at small angular scales. 
We expect significant non-Gaussianity in the tSZ, CIB and radio galaxy emission. 

While we have masked the brightest of these sources, there remains a background of dimmer objects.
With the QE, the lensing estimate can be further ``hardened" against bias from  dim point sources if the foreground trispectrum is known \cite{osborne_extragalactic_2014}, or it can be made insensitive to sources with specific profiles \cite{sailer_lower_2020}. 
While an analogous approach is possible in the case of polarization for a Bayesian estimator by marginalizing over the foregrounds using a Poissonian prior, this is in general not possible for temperature, given the more complex distribution of the foregrounds \cite{qu_impact_2024}.

In the implementation of the Bayesian lensing estimator used in this work (see \S\ref{sec:muse}), we assume the non-CMB signals in the ILC maps can be modeled as a Gaussian realization of the total ILC residual power spectrum. 
While this should be a reasonable assumption for instrumental noise, we expect astrophysical foreground emission to show non-Gaussianity. 
We note that while the ILC maps will minimize the total non-CMB variance in the maps, this may in fact enhance the contribution from non-Gaussian signals. 
A Bayesian lensing estimator using a forward model at the field level can easily be implemented on an ILC map, so the goal of this work is to establish the extent to which neglecting the non-Gaussian nature of foreground signals biases the power spectra recovered with such a method.


\section{Results}
\label{sec:Results}

In the following section we describe the performance of MUSE for three cases varying the multipoles used, or the input foreground simulations. \new{We use the same set of 13 independent, lensed CMB skies in all cases in order to separate the effect of these modeling decisions from the sample variance of the CMB. The lensed CMB skies are extracted from the Agora simulations.  The three cases are as follows}

\begin{enumerate}
  \item[]\textbf{I}. \new{We use non-Gaussian foregrounds from the Agora simulations when constructing the ILC maps. 
  The foreground emission is correlated with the lensing potential $\phi$ map.
  The low-pass filter in Eqn (\ref{eqn:data_model}) is set at $\ell_{\rm{max}}=3500$.} 
    \item[]\textbf{II}. \new{The same non-Gaussian foregrounds as \textbf{I}, but the low-pass filter is set at $\ell_{max}=3000$. }
     \item[]\textbf{III}. \new{This uses the same low-pass filter at $\ell_{max}=3000$ as \textbf{II}, however, matching Gaussian foreground power is used instead of the non-Gaussian foregrounds from Agora. The Gaussian foreground power is uncorrelated with the lensing potential $\phi$ map. }
\end{enumerate}

We model the residual, non-CMB power (hereafter foregrounds) in the data maps as Gaussian distributed.
The output bandpowers,  $\bar{\theta}_{(i)}$,  are the average of the bandpowers inferred by MUSE from the 13 maps for each $i \in$ \textbf{I-III}. To assess the level of bias on the $\bar{\theta}_{(i)}$, we require its covariance. 
The formally correct covariance ($\bar{\Sigma}_{(i)}$) to use here would be the average of the covariances over the thirteen independent maps, with an extra factor of $\frac{1}{13}$. But as the covariance for the $j$th MUSE run for the $i$th case ($\Sigma_{j(i)}$) as defined in Eqn (\ref{eqn:muse_covariance}) is calculated using the forward model in Eqn (\ref{eqn:data_model}), the only difference between these $\Sigma_{j(i)}$, is the source masking on the thirteen maps. 
Given this, we use the approximation $\bar{\Sigma}_{(i)} \simeq \frac{1}{13} \Sigma_{j=1(i)}$. 
As the forward model of cases \textbf{II} and \textbf{III} are the same, this leaves only two $\Sigma$ to calculate via the $J$ and $H$ matrices in Eqn (\ref{eqn:muse_covariance}). We use 500 simulations to calculate $J$ in the final covariance,  and perform linear shrinkage on a combination of $J$ and $H$ before the covariance is calculated as described in appendix C of Ref.\cite{ge_cosmology_2024}. 

\subsection{Bias estimation}
\label{sec:fg_power_estimation}
\new{ We test for bias in two ways. 
 First, following equation (8) in \cite{qu_impact_2024}, we
estimate the bias on the recovered lensing potential amplitude, $A^{\phi\phi}$, normalized to units of the expected error on this quantity from a lensing analysis of  two years of the SPT-3G Main survey, $\frac{\Delta A^{\phi\phi}}{\sigma}$:}

\begin{equation}\label{eqn:bias_systematic_ovr_stat}
    \frac{\Delta A^{\phi\phi}}{\sigma} =  \Big{(}  \Sigma_{b} \frac{\langle A_b^{\phi\phi}\rangle-1}{\hat{\sigma}_b(A^{\phi\phi})^2} \Big{/} \Sigma_b\frac{1}{\hat{\sigma}_b(A^{\phi\phi})^2} \Big{)} \sqrt{\Sigma_b \frac{1}{\sigma_{b,\rm{spt}}(A^{\phi\phi})^2}},
\end{equation}
\new{ where $\hat{\sigma}_b(A^{\phi\phi}) = \sqrt{\rm{diag}(\bar{\Sigma}_{(i)})
 }$ is the standard error on the mean in bin $b$, and $\sigma_{b,\rm{spt}}$ is the expected error bar assuming this analysis were carried out on one map matching the 1500 square degree SPT-3G main field.}
 
 \new{Second, we test for the existence of a bias by calculating a } Probability To Exceed (PTE) and an equivalent Gaussian $\sigma$ for these PTE, given \new{the mean recovered power and } the underlying truth of Agora. Specifically, for a case in \textbf{I-III} indexed by $i$, we form a $\chi_{(i)}
^2$ as:

\begin{equation}\label{eqn:chi_squared}
    \chi_{(i)}^2 = (\bar{\theta}_{(i)} - \theta^*)^{\rm{T}} \bar{\Sigma}_{(i)}^{-1} (\bar{\theta}_{(i)}- \theta^*),
\end{equation}
where $\theta^*$ are the true underlying bandpower amplitudes of the Agora simulation. 
\new{As the $J$ matrix in equation (\ref{eqn:muse_covariance}) is calculated from 500 simulations, we assume the above follows an F-distribution}, and compute the PTE and equivalent Gaussian $\sigma$'s. 

\new{The results of both tests are summarized in Tables \ref{Table:delta_Aphi} and \ref{Table:ptes_and_sigmas}, and discussed in \S\ref{sec:bandpowers}.}

\subsection{Foreground Power Estimation}
\label{sec:fg_power_estimation}
\new{In cases \textbf{I} and \textbf{II},  we estimate the foreground power  directly from each ILC map. We fix $C_{\ell}^{TT}$ and $C_{L}^{\phi\phi}$ to their fiducial values and fit for a flat foreground power spectrum in the multipole range $2000 \textless \ell  \textless 4000 $.
We use this initial estimate as the starting point for MUSE and perform one MUSE step  (steps 1 and 2 in \ref{sec:muse})
    to estimate the binned foreground spectrum (\{$C_{\ell,\rm{fid}}^{gg}$\}), for each of the 13 ILC maps in cases \textbf{I} or \textbf{II}. 
The recovered foreground power spectrum from case \textbf{II} is also used for case \textbf{III}.
One MUSE step produces a reasonable starting point for the full analysis, as one step is sufficient to recover the total power in each bin to within 5\% of the data. }

With the \{$C_{\ell,\rm{fid}}^{gg}$\} in hand, we free all three bandpower amplitudes, ($\theta = A_b^{\phi\phi}, A_b^{TT}, A_b^{gg}$), and let $\theta$ converge to the value that satisfies Eqn (\ref{eqn:muse_score_equation}).
\new{We have compared the final foreground bandpower estimates to the true full-sky foreground power spectrum in the ILC maps, and find them to be in good agreement. }

\subsection{Bandpowers}
\label{sec:bandpowers}

For each MUSE inference on an individual map (hereafter MUSE run) we use 200 simulations to compute the expectation value on the RHS of Eqn (\ref{eqn:muse_score_equation}), giving an expected MonteCarlo error on the output $\hat{\theta}$ inference of an individual map of $\lesssim$ 0.07$\sigma$.
To carry out the maximization of $T$, $\phi$, and $g$ with respect to log$\mathcal{P}$, we use the alternating coordinate descent/pseudo Newton-Raphson algorithm from Ref.\cite{millea_bayesian_2019}.
We set the minimum temperature mode included to $\ell_{\rm{min}}^T$=200, and use a binning of $\Delta \ell$ = 50 for $C_{\ell}^{TT}$. For the lensing power spectrum inference we use 12 logarithmically spaced bins in 100 \textless $L$ \textless 3000. Given that the $\ell_{\rm{max}}$=3000 runs contain very little information  on $C_{L}^{\phi \phi}$ in the final 2300 $ \lesssim L$ \textless  3000 bin, we find this bin to be poorly constrained, and so excise it. Note that excising a parameter from a converged MUSE inference constitutes marginalizing over that parameter.
It is also worth noting that both cases \textbf{I} and \textbf{II} use the same input data patches, and the same sets of 200 simulation seeds to compute the MUSE estimate, so the only substantive difference between these cases is $\ell_{\rm{max}}$.

For the foreground spectrum binning, we use a large initial bin with 200 $\textless \ell \textless $ 1800. We choose these bin edges such that the S/N for the foregrounds is greater than 5. For all subsequent bins we use $\Delta \ell$ = 200. 
As an estimate of the foreground signal (S), we use the foreground power estimate from data described above, and for the noise with respect to the foreground signal (N), we use the fiducial lensed temperature and noise power.
We use cutouts of the point source mask described in \ref{sec:Agora}, and a 1 degree apodization mask around the edge of the maps for all three cases.

\begin{figure}
    \centering
    \includegraphics[width=1.0\linewidth]{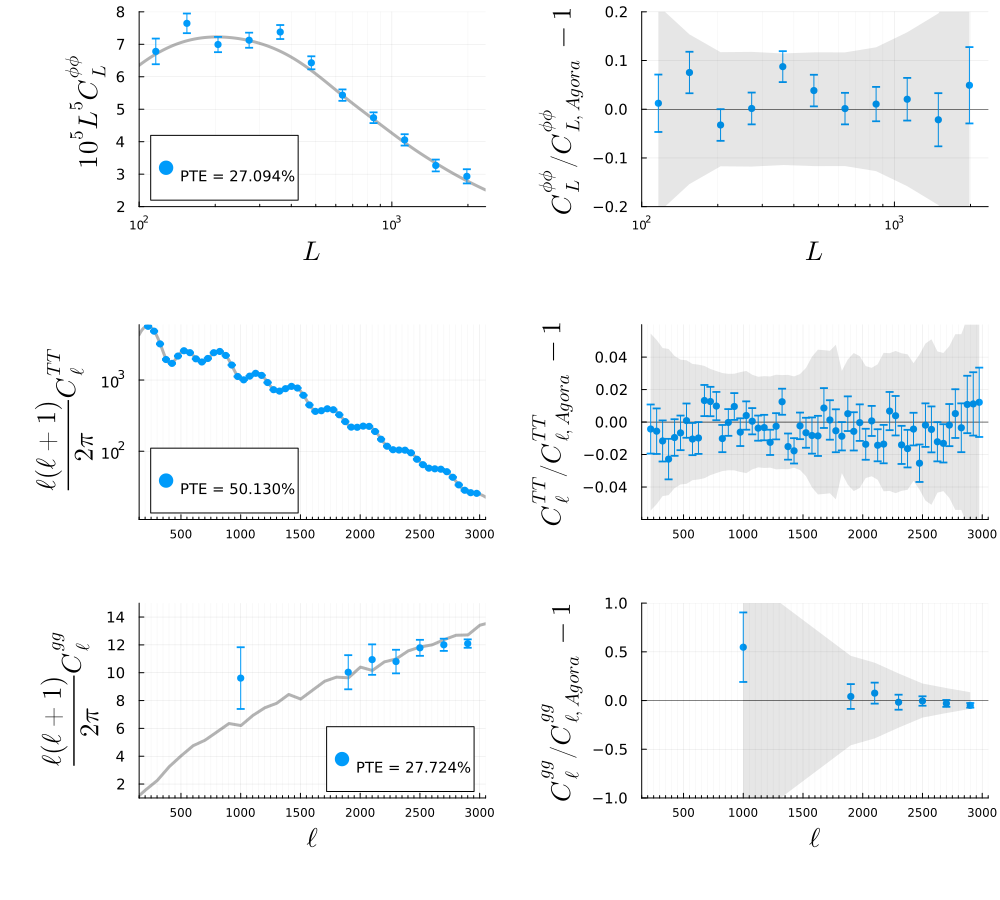}
    \caption{Average results over an ensemble of MUSE runs carried out on 13 independent maps matching case \textbf{III} (Gaussian foregrounds, uncorrelated with $\phi$, and $\ell_{\rm{max}}$=3000). 
    \textit{Left:} Absolute estimate of the power for $\phi$, $T$, and $g$ from top to bottom. The error bars are the standard error on the mean ($\hat{\sigma}$), the grey lines are the true power in the Agora simulations, and the legends include reduced $\chi^2$'s from Eqn (\ref{eqn:chi_squared}). \textit{Right:} The fractional difference between the estimated power and the true underlying power in the Agora simulations, with error bars as above, and the grey shaded area the scatter from a single MUSE run ($\sigma_{\rm{spt}}$). All bandpowers are recovered at less than 3$\hat{\sigma}$ from the truth. PTE and equivalent $\sigma$'s for joint inference of the different bandpowers are in Table \ref{Table:ptes_and_sigmas}.}
    \label{fig:gauss_fg_pwr_and_bpwrs}
\end{figure}

\subsubsection{Gaussian Foregrounds - Case III}
\label{sec:gauss_bandpowers}
We begin with case \textbf{III} (Gaussian foregrounds) and use it as a baseline test on the inference pipeline, as it has input data that match our forward model as defined in Eqn (\ref{eqn:data_model}), barring the projection from full-sky.
From top to bottom, Fig \ref{fig:gauss_fg_pwr_and_bpwrs} shows the averaged results over thirteen runs for $C_L^{\phi\phi}, C_{\ell}^{TT}$, and $C_{\ell}^{gg}$ for case \textbf{III}.
The left panels show the absolute power, and the right panels show the fractional difference between the estimated power and the true underlying power in the Agora simulations, with all error bars being the standard error on the mean as described in \S \ref{sec:fg_power_estimation}. The shaded area in the right panels is the scatter from one MUSE run, $\sigma_{\rm{spt}} = \sqrt{13}\hat{\sigma}$.
We see no significant bias to the output $C_L^{\phi\phi}, C_{\ell}^{TT}$, or $C_{\ell}^{gg}$ for case \textbf{III}, with all recovered bandpowers less than 3$\hat{\sigma}$ from truth.

\new{We do not detect a statistically significant bias in the recovered lensing potential amplitude with Gaussian foregrounds, finding $\frac{\Delta A^{\phi\phi}}{\sigma} = (0.5\pm 0.3)\sigma_{\rm{spt}}$. 
The PTE test is also consistent with no bias, with a PTE=27\% for the lensing potential amplitude and similarly high PTE for the other measured quantities. 
Future work is necessary to test the possible existence of lower levels of bias. }

\new{Curiously, although the PTE for each individual term is  $>26$\%,} the joint inference of $A^{\phi\phi}$, $A^{TT}$ and $A^{gg}$ is 0.53\%. 
In fact the joint inference of $A^{TT}$ alone with $A^{gg}$ is 0.3\%, so it is this tension that drives this apparent bias. 
As the focus of this work is on the unbiased recovery of $A^{\phi\phi}$ and $A^{TT}$, we do not investigate this apparent tension further.

\subsubsection{Non-Gaussian Foregrounds}
\label{sec:nongauss_bandpowers}

\begin{figure}
    \centering
    \includegraphics[width=1.0\linewidth]{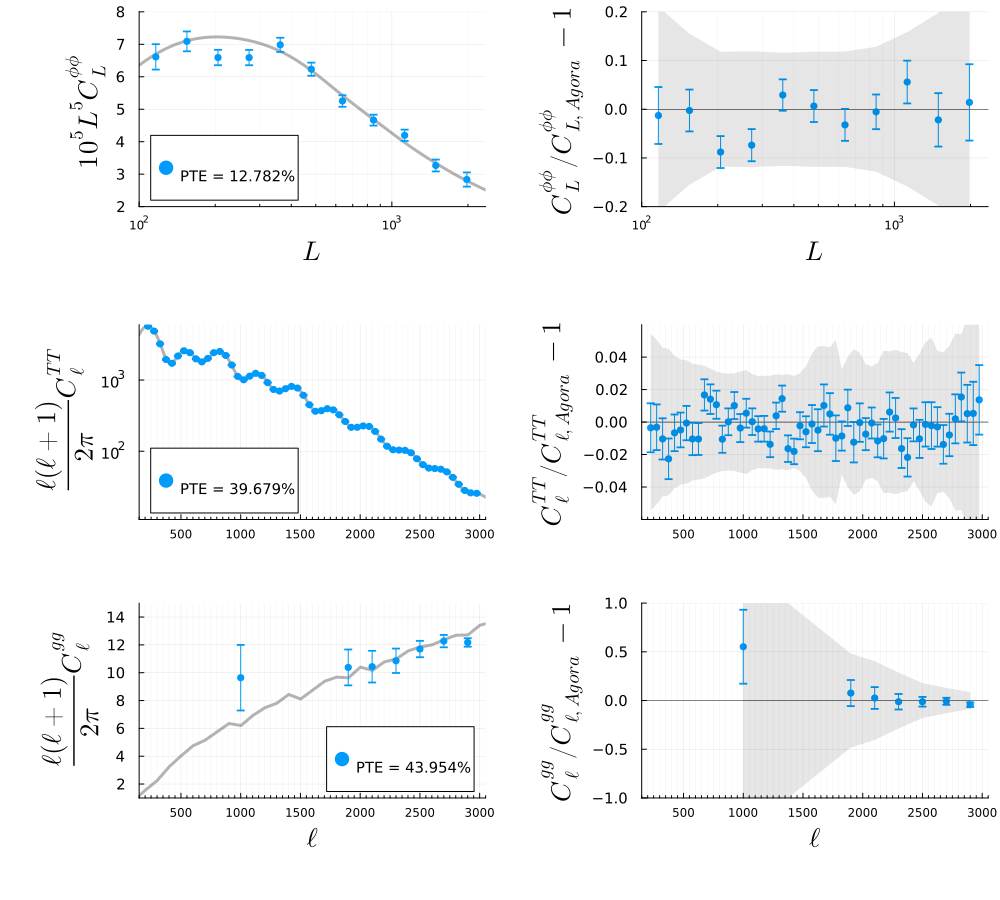}
    \caption{As in Fig \ref{fig:gauss_fg_pwr_and_bpwrs}, but for input data matching case \textbf{II} (non-Gaussian foregrounds, correlated with $\phi$, and $\ell_{\rm{max}}$=3000). No significant bias is apparent, with the greatest discrepency being the third and fourth $A^{\phi\phi}_b$ recovered at 2.7$\hat{\sigma}$ and 2.2$\hat{\sigma}$ respectively. All $A^{\phi\phi}_b$ are recovered at less than 3$\hat{\sigma}$. PTE and equivalent $\sigma$'s for joint inference of the different bandpowers are in Table \ref{Table:ptes_and_sigmas}.}
    \label{fig:NG_fg_pwr_and_bpwrs_lmax3k}
\end{figure}

\textbf{Case II: $\ell_{max}=3000$:} \new{We present the results for the non-Gaussian foregrounds of case \textbf{II} in Fig. \ref{fig:NG_fg_pwr_and_bpwrs_lmax3k}.}
The recovery of the $A^{TT}$ and $A^{gg}$ bandpowers is almost identical to \new{the Gaussian case \textbf{III}, and all $A_{b}^{\phi\phi}$ bandpowers recovered to within 3$\hat{\sigma}$ of the underlying truth.} However, we see a systematic shift downwards in the $A_{b}^{\phi\phi}$ for $L \lesssim $ 400 \new{compared to the Gaussian foregrounds of} case \textbf{III}. In particular, the third and fourth bandpowers ($L = $205, 272) shift to 2.7$\hat{\sigma}$ and 2.2$\hat{\sigma}$ from the underlying truth. 

\new{We do not detect a statistically significant bias in the recovered lensing potential amplitude with non-Gaussian foregrounds at $\ell_{max}=3000$, finding $\frac{\Delta A^{\phi\phi}}{\sigma} = (-0.4\pm 0.3)\sigma_{\rm{spt}}$. 
The PTE test is also consistent with no bias, with a PTE=13\% for the lensing potential amplitude and higher PTE for the other measured quantities. 
As in the Gaussian case, future work is necessary to test the possible existence of lower levels of bias. }

\new{Mirroring the Gaussian case, the PTE for joint inference of all terms ($A^{\phi\phi}$, $A^{TT}$, $A^{gg}$) is low even though the PTE of each individual term is high. 
This is again due to the interplay between $A^{gg}$ and $A^{TT}$. }


\begin{figure}
    \centering
    \includegraphics[width=1.0\linewidth]{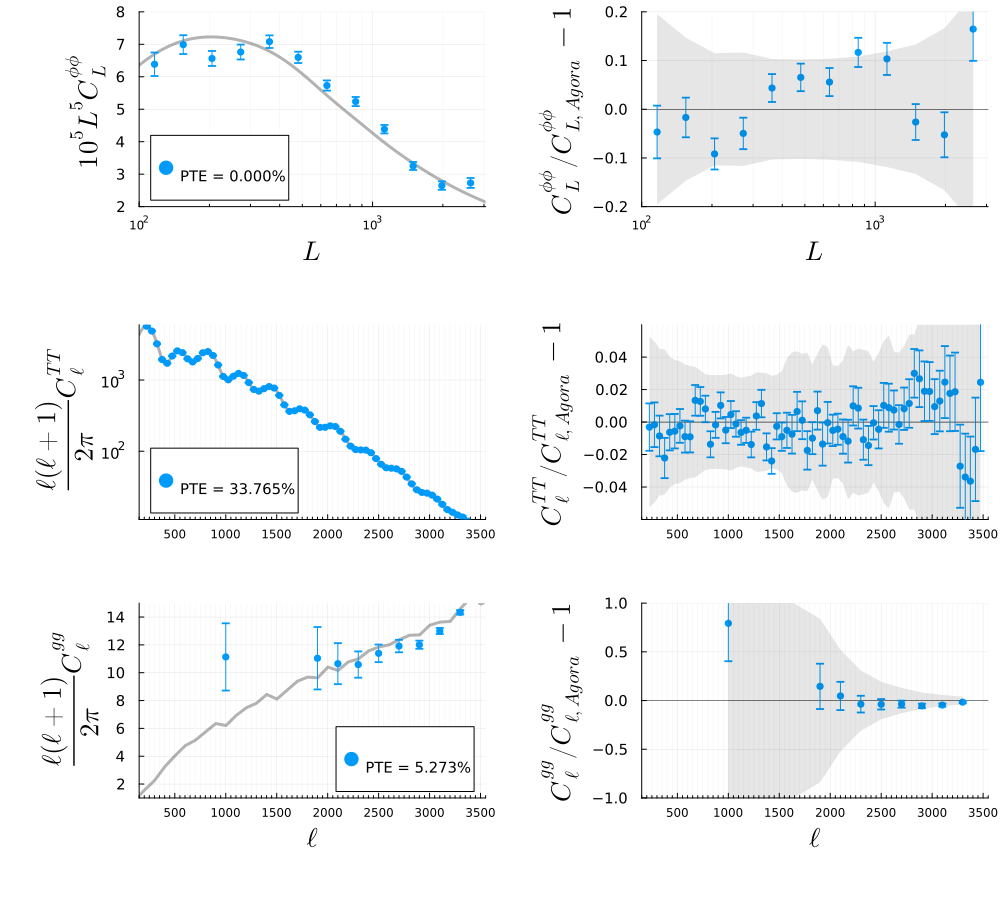}
    \caption{As in Fig \ref{fig:gauss_fg_pwr_and_bpwrs}, but for input data matching case \textbf{I} (non-Gaussian foregrounds, correlated with $\phi$, and $\ell_{\rm{max}}$=3500). We note that the $A_b^{TT}$ and $A_b^{gg}$ are recovered without bias, as reflected by the PTE in Table \ref{Table:ptes_and_sigmas}. However, the $A_b^{\phi\phi}$ show a bias with a PTE equivalent to that of a 5.3$\sigma$ Gaussian fluctuation for this higher $\ell_{\rm{max}}$ case.}
    \label{fig:NG_fg_pwr_and_bpwrs_lmax3500}
\end{figure}


\begin{table}[h]
    \centering
    \caption{\new{Bias to the lensing inference ($\frac{\Delta A^{\phi\phi}}{\sigma}$) using equation (\ref{eqn:bias_systematic_ovr_stat}) with its uncertainty. Cases \textbf{I-III}, differ only in the nature of the foregrounds and $\ell_{\rm{max}}$ as described in the text and noted in the first column. The $\frac{\Delta A^{\phi\phi}}{\sigma}$ column shows the 13 result average for each case lacks the statistical power to rule out a $\lesssim 0.3\sigma_{\rm{spt}}$ bias. It does, however, show the bias is slightly higher for case \textbf{I}, indicating a greater bias when modes up to $\ell_{\rm{max}}$=3500 are included in the inference.}
    }
    \begin{tabular}{c c c | c }
        \hline
        \hline
         & & $\ell_{\rm{max}}$ & $\frac{\Delta A^{\phi \phi}}{\sigma}$ \\
        \hline
        \textbf{I} & NG & 3500 & $0.7\pm0.3$  \\ 
        
        \textbf{II} & NG & 3000 & $-0.4\pm0.3$   \\ 
        \hline
        \textbf{III} & G & 3000 & $0.5\pm0.3$ \\ 
        \hline
        \hline
    \end{tabular}
    \label{Table:delta_Aphi}
\end{table}

\begin{table}[h]
    \centering
    \caption{PTE in percent with equivalent gaussian $\sigma$'s in parentheses \new{assuming equation (\ref{eqn:chi_squared}) follows an F-distribution.} Cases \textbf{I-III}, differ only in the nature of the foregrounds and $\ell_{\rm{max}}$ as described in the text and noted in the first column. Bandpower amplitude inferences of individual fields ($A^{xx}$) and combinations are shown. The PTE  show a significant bias to the $A^{\phi\phi}$ inference for case \textbf{I}. Note the similarity between cases \textbf{II} and \textbf{III} in their PTE and equivalent $\sigma$'s, in contrast with case \textbf{I}. In particular, the 5.3$\sigma$ discrepancy of the $A^{\phi\phi}$ PTE, indicating a bias when modes up to $\ell_{\rm{max}}$=3500 are included in the inference.
    }
    \begin{tabular}{c c c | c c c c c }
        \hline
        \hline
         & & $\ell_{\rm{max}}$ &  $A^{\phi \phi}$  & $A^{TT}$  & $A^{gg}$ & $A^{\phi \phi}$ \& $A^{TT}$ & All\\
        \hline
        \textbf{I} & NG & 3500  & $4\times10^{-6}$ (5.3) & 33.76 (0.4)  & 5.27 (1.6) &  0.22  (2.9) & $4\times10^{-4}$  (4.5) \\ 
        
        \textbf{II} & NG & 3000 & 12.78  (1.1) & 39.68 (0.3) & 43.95  (0.2)  & 25.25  (0.7) & 0.14  (3.0)  \\ 
        \hline
        \textbf{III} & G & 3000  &  32.47 (0.5) & 45.15 (0.1) &  26.24 (0.6)  &  53.07 (0.1) &  0.53 (2.6) \\ 
        \hline
        \hline
    \end{tabular}
    \label{Table:ptes_and_sigmas}
\end{table}

\noindent\textbf{Case I ($\ell_{max}=3500$):} 
\new{We present the results when including angular multipoles out to 3500 in Fig. \ref{fig:NG_fg_pwr_and_bpwrs_lmax3500}.}
 We see the same behaviour in $A_{b}^{\phi\phi}$ for $L \lesssim $ 400, and a further upward shift of the seventh, eighth, and ninth bins ($L = $ 637, 846, and 1124 respectively) not seen in case \textbf{II}. The shifts result in an offset from the underlying truth of Agora by 1.9$\hat{\sigma}$, 3.9$\hat{\sigma}$ and 3.1$\hat{\sigma}$ respectively, compared to 1.0$\hat{\sigma}$, 0.1$\hat{\sigma}$ and 1.2$\hat{\sigma}$ for the same bins in case \textbf{II}.
It is worth reiterating here that cases \textbf{I} and \textbf{II} use the same input data, and the same sets of 200 simulations per step to compute the MUSE estimate. So this shift is entirely due to the inclusion of small scale modes in the analysis.

\new{We see a bias of $\frac{\Delta A^{\phi \phi}}{\sigma} = (0.7\pm 0.3)\sigma_{\rm{spt}}$ in the recovered lensing potential amplitude due to the inclusion of these smaller angular scale modes. 
The existence of this bias on the lensing potential is confirmed by the PTE test, where we find a PTE of $4\times10^{-6}$\%, indicating a highly significant disagreement between the input and recovered lensing potential. }

\begin{figure}
    \centering
    \includegraphics[width=0.85\linewidth]{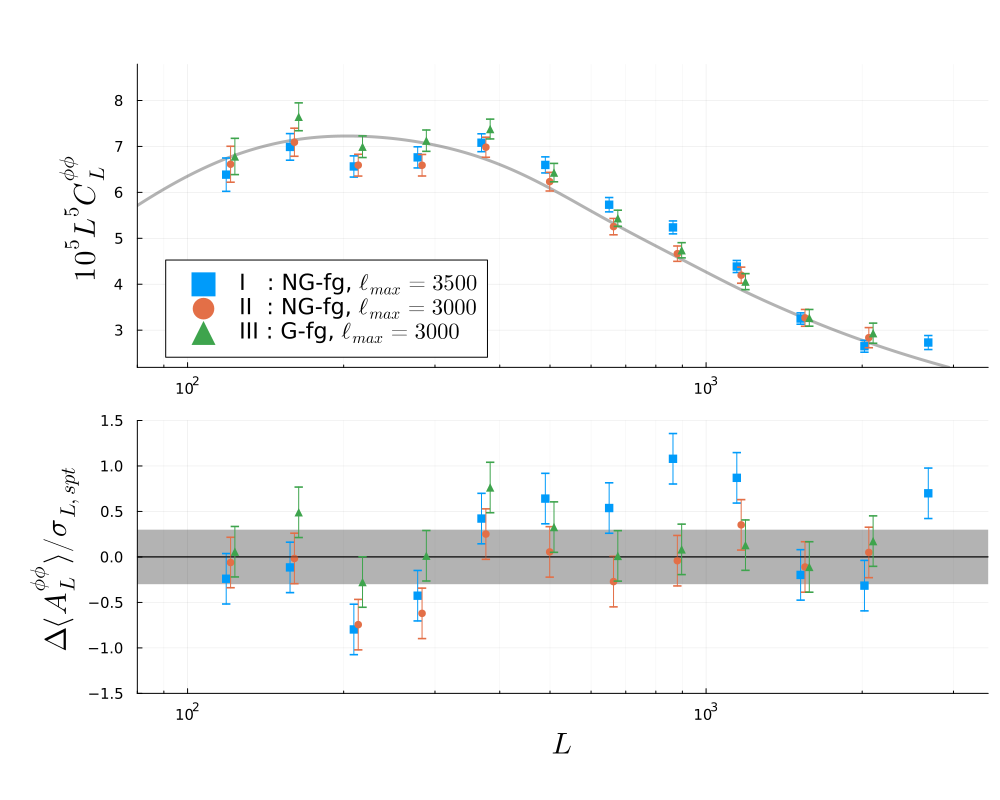}
    \caption{Averaged lensing bandpower results from an ensemble of MUSE runs carried out on 13 independent data-patches for cases \textbf{I} (blue square), \textbf{II} 
 (orange circle), and \textbf{III} (green triangle) as described in the text, offset in $L$ for clarity. \textit{Top:} Absolute estimate of the power, where the error bars are the standard error on the mean ($\hat{\sigma}$). \textit{Bottom:} The fractional difference between the estimated power and the true underlying power in the Agora simulations in units of uncertainty from a single MUSE run ($\sigma_{\rm{spt}}=\sqrt{13}\hat{\sigma}$). The grey shaded area covers $\pm 0.3\sigma_{\rm{spt}}$. Note that the eighth, and ninth bandpowers ($L$ = 846, and 1123) shift upwards to  1.1$\sigma_{\rm{spt}}$, and 0.9$\sigma_{\rm{spt}}$ away from the Agora $C_L^{\phi\phi}$. This is not detected in case \textbf{II} (non-Gaussian foregrounds, $\ell_{\rm{max}}$=3000), where the same bandpowers are 0.04$\sigma_{\rm{spt}}$, and 0.4$\sigma_{\rm{spt}}$ away from the Agora $C_L^{\phi\phi}$.
    Table \ref{Table:ptes_and_sigmas} shows that this shift in the $A^{\phi\phi}_b$ gives a PTE equivalent to that of  5.3$\sigma$ Gaussian fluctuation. As case \textbf{I} and case \textbf{II} are identical, barring their $\ell_{\rm{max}}$, we attribute these shifts as being sourced by the non-Gaussianity of the smaller scales included in case \textbf{I}.}
    \label{fig:Aphi_bpwrs}
\end{figure}

Fig \ref{fig:Aphi_bpwrs} shows the $A_{b}^{\phi\phi}$ for cases \textbf{I-III}, where this shift is more readily apparent. The lower panel of Fig \ref{fig:Aphi_bpwrs} differs from the previous, where we show the averaged bias to $A^{\phi\phi}$ normalised to $\sigma_{\rm{spt}}$ in each bin $b$.
\begin{equation}\label{eqn:avg_Aphi_bias_wrt_spt2yr}
    \Delta \langle A_b^{\phi\phi} \rangle/\sigma_{b,\rm{spt}}=  \frac{\langle A_b^{\phi\phi}\rangle-1}{\sigma_{b,\rm{spt}}(A^{\phi\phi})} 
\end{equation}
We also include a 0.3$\sigma_{\rm{spt}}$ contour on the lower panel as the grey shaded area.
\new{Note that the bins for case  \textbf{II} match \textbf{III} closely, barring the third and fourth bandpowers, indicating there may be some bias present for case \textbf{II}.
However, only the third, fourth and ninth bandpowers lie outside of the 0.3$\sigma_{\rm{spt}}$ contour for case \textbf{II}, whereas case \textbf{I} performs much more poorly with only three of the twelve bandpowers lying within 0.3$\sigma_{\rm{spt}}$ of Agora.}

\new{We interpret the upward shift of the high $L$ bins in case \textbf{I} as the lensing model mistaking the non-Gaussianity of the foreground distribution for lensing induced non-Gaussianity and biasing the $A^{\phi\phi}$ inference. 
Due to these results, we do not recommend using MUSE with temperature data above $\ell=3000$ unless further steps are taken to mitigate the non-Gaussianity of the foregrounds.}

\subsection{Covariance Matrices}
\begin{figure}
    \centering
    \includegraphics[width=1.\linewidth]{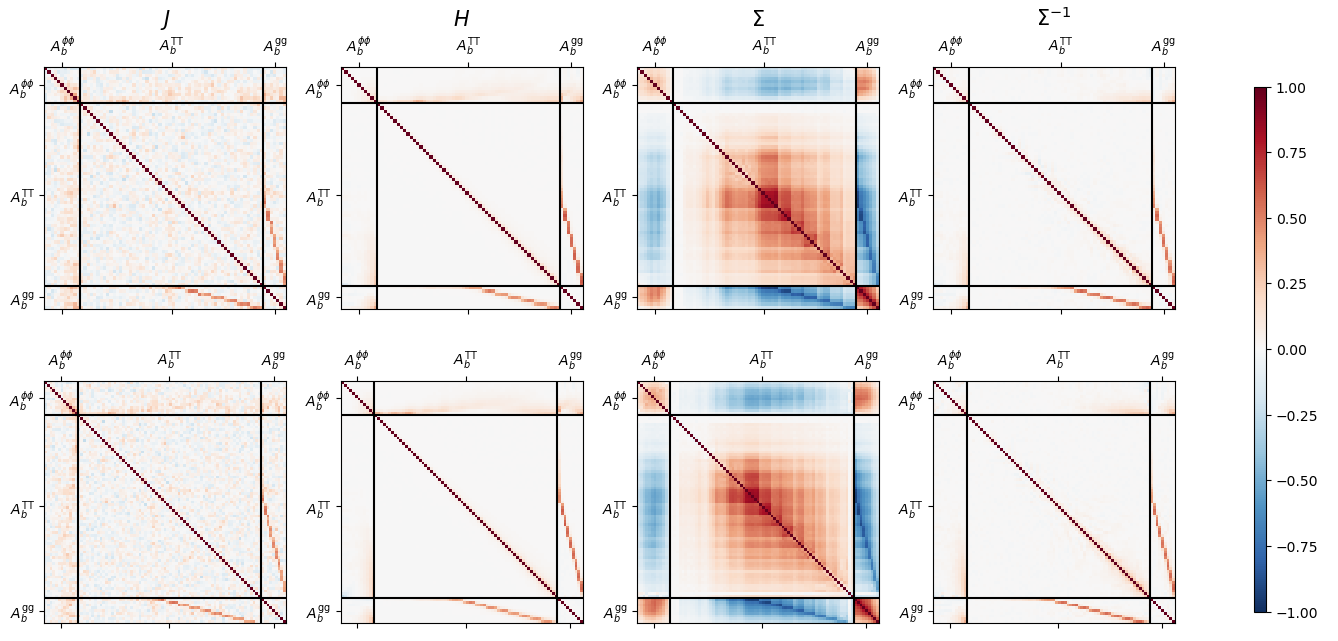 }
    \caption{Typical covariances ($J,H,\Sigma, \Sigma^{-1}$) plotted as correlation matrices for a MUSE run on a single ILC map. Linear shrinkage is performed in computing $\Sigma$ following appendix C of Ge et al \cite{ge_cosmology_2024}.  \textit{Top}: $\ell_{\rm{max}}$=3000. \textit{Bottom}: $\ell_{\rm{max}}$=3500}
    \label{fig:lmax3kand3500_covs}
\end{figure}

We show in Figure \ref{fig:lmax3kand3500_covs} a set of MUSE matrices from a single MUSE run for $\ell_{\rm{max}}^T=3000$ (cases \textbf{II} and \textbf{III}) in the top panels, and 3500 (case \textbf{I}) on the bottom.
Of particular note in all of them is the off-diagonal correlation structure in the $A^{TT} \times A^{gg}$ blocks, where the correlation is between parameters in the same bin. 
Zooming in you can see there are 4 pixels along $T$ ($\Delta\ell^T=50$) for one in $g$ ($\Delta\ell^g=200$ for $\ell >$  1800). 
For J, the covariance of the score at the MAP, this off-diagonal structure shows a positive correlation between the variance of the score with respect to $A^{TT}$, and $A^{gg}$.
Essentially, a variation in power due to $T$ can approximately be fit to the model by a commensurate variation in $g$.
This is as one might expect, given that the main distinction between power in $T$ and $g$ in the ILC data model (Eqn (\ref{eqn:data_model})) is that $T$ is lensed.

The H-matrix has similar, yet asymmetric correlation. This asymmetry is due to the H matrix being formed from a further derivative of the score at the MAP that only operates on the parameters controlling the distribution of the data. 
As suggested in Ref.\cite{millea_muse_2022}, one can think of an entry in H, as how the score at the MAP responds, on average, to infinitesimal shifts in the parameters controlling the distribution of the data. 
In the vertical $A^{TT} \times A^{gg}$ block we see a stronger correlation than in the horizontal. The vertical block corresponds to injected power in $T$, which has a greater relative effect on the score with respect to $g$ up to this $\ell \leq 3000$, given the power level in $T$ is  still higher than that of $g$ in the ILC map.
We also see some correlation in the $A^{\phi\phi} \times A^{TT}$ blocks, with the largest effect in the top horizontal block, similar to what Ref.\cite{millea_muse_2022} found in polarization. This corresponds to the score at the MAP responding to injected power in $A^{\phi\phi}$ by increasing power in $A^{TT}$. 

Most of the interesting off-diagonal features in the MUSE covariance, $\rm{Cov}(A^{XX}, A^{YY})=\Sigma$, are sourced from the $A^{TT} \times A^{gg}$ correlation present in $J$ and $H$.
The $A^{TT} \times A^{gg}$ blocks of $\Sigma$ show an anti-correlation in the scatter of these bandpower amplitudes, for the same reason that the covariance of the score at the MAP ($J$) showed a correlation, i.e. a higher $A^{TT}$ to explain the variance in the data favours a lower $A^{gg}$, and vice-versa. 
This correlation between $A^{TT}$ and $A^{gg}$ causes some correlation between $A^{TT}_b$  bandpowers that share the same $A^{gg}_b$ bin.
However, the correlation between $A^{TT}$'s extends beyond those sharing a single $A^{gg}_b$, as it is \textit{lensed} $T$ and $g$ that have equivalent effects on the total power in the data.
So given fixed total power in a data realization, one could have a decrement in foreground power, balanced by an increase in lensed power, which would require more power in unlensed temperature over a range of bins, determined by the mode-coupling effect of $\phi$. 
Conversely, a small decrement in foreground power could also be compensated for by less lensing, which is why we see positive correlation in the $A^{\phi\phi} \times A^{gg}$ block.

\newpage

\section{Conclusion}
\label{sec:Conclusions}

\new{In this paper we demonstrate reconstruction of the lensing potential} and temperature power spectra when modeling non-Gaussian foreground power in a MV-ILC map as a simple Gaussian distributed field.
We carry out the joint inference with a Bayesian inference method known as MUSE \cite{millea_muse_2022}, on a set of thirteen MV-ILC maps made using the Agora suite of correlated extragalactic simulations \cite{omori_agora_2024}. The assumed noise levels, listed in Table  \ref{Table:map_dimensions_and_noise_levels}, approximately match SPT-3G at 2 years of integration time \cite{prabhu_testing_2024}. 

\new{We find no evidence for bias in the MUSE recovery with Gaussian and non-Gaussian foregrounds when limited to a maximum angular multipole of $\ell = 3000$, although more work is necessary to rule out the possibility of low levels of bias. 
We detect a statistically significant bias is induced by non-Gaussian foregrounds when using angular multipoles up to $\ell = 3500$. 
The magnitude of the bias is $(0.7\pm 0.3)\sigma$ for a survey like the SPT-3G Main survey. 
We explain the bias in the  MUSE estimate as the lensing model attributing the non-Gaussianity in the foreground emission to the induced non-Gaussianity of lensing on the CMB.}




We discuss the correlation structure observed in both the covariance matrix, and the $J$ and $H$ matrices.
We attribute the significant off-diagonal correlations of the covariance matrix to the partial degeneracy in the lensing model between lensed temperature power and foreground power. In principle, one might reduce the degeneracy by using multi-frequency maps instead of an ILC map to leverage the different SEDs of the CMB and foreground components.  
However, this would significantly increase the computational cost due to the requirement to marginalize over more parameters.

Future work should explore adaptations of MUSE to include more information (i.e. increase $\ell_{\rm{max}}$), such as downweighting the non-Gaussian contribution of the foregrounds by using a needlet basis \cite{remazeilles_cmb_2011, surrao_constraining_2024-1}, or using a cross-ILC method \cite{raghunathan_cross-internal_2023}. Alternatively, one could attempt to directly model the non-Gaussianity of the foregrounds.
To do the latter at the field level however, would require the ability to forward model the correlated, non-Gaussian foregrounds, \new{which may be feasible using generative machine learning techniques \cite{prabhu_learning_2025, 2025arXiv250521220M}.}



\acknowledgments

Melbourne authors acknowledge support from the Australian Research Council’s Discovery
Projects scheme (DP210102386). 
MD acknowledges support from the University of Melbourne Research Scholarship.
SR acknowledges support by the Illinois Survey Science Fellowship from the Center for AstroPhysical Surveys at the National Center for Supercomputing Applications.
LK acknowledges support from Michael and Ester Vaida. 
The SLAC authors acknowledge support by the Department of Energy, under contract DE-AC02-76SF00515.
WLKW acknowledges support from an Early Career Research Award of the Department of Energy and a Laboratory Directed Research and Development program as part of the Panofsky Fellowship program at the SLAC National Accelerator Laboratory.
This research was supported by The University of Melbourne’s Research Computing Services and the Petascale Campus Initiative.



\newpage
\bibliographystyle{apsrev4-1} 
\bibliography{Paper1_refs}
\end{document}